\documentclass[aps,prl,reprint,amsmath,amssymb]{revtex4-2}
\usepackage{graphicx}
\usepackage{bm}
\usepackage[colorlinks,allcolors=blue]{hyperref}

\usepackage{slashed}
\usepackage{amsmath}
\usepackage{amssymb}
\usepackage{amsthm}
\usepackage[dvipsnames]{xcolor}
\usepackage{indentfirst}
\usepackage{psfrag}
\usepackage{cleveref}
\Crefname{equation}{Eq.}{Eqs.}
\usepackage[utf8]{inputenc}

\hypersetup{colorlinks=true, linkcolor=blue, urlcolor=blue, citecolor=blue, linktocpage=true}

\renewcommand{\l}{\lambda}

\newcommand\e{\text{e}}
\renewcommand{\d}{\partial}
\newcommand{\tg}{\tilde{g}}

\newcommand{\BL}{B-L}
\newcommand{\Lem}{L_e-L_\mu}
\newcommand{\Let}{L_e-L_\tau}
\newcommand{\Lmt}{L_\mu-L_\tau}
\newcommand{\LLo}{L_e-L_{\mu/\tau}}
\newcommand{\mV}{m_{V,\rm{eff}}}
\newcommand{\mVbar}{\bar{m}_{V,\rm{eff}}}
\newcommand{\mF}{m_{\phi,\rm{eff}}}
\newcommand{\geff}{g_{\rm eff}}

\begin{document}

\title{Screening of New Abelian Long-Range Forces}

\author{Marco Costa$^{1,2}$}
\email{marcocosta@kitp.ucsb.edu}
\author{Andrey Shkerin$^1$}
\email{ashkerin@perimeterinstitute.ca}
\author{Ananda Thomlinson$^{3}$}
\email{anandagt@student.ubc.ca}

\affiliation{$^1$Perimeter Institute for Theoretical Physics, 31 Caroline St N, Waterloo, ON N2L 2Y5, Canada}
\affiliation{$^2$Kavli Institute for Theoretical Physics, Santa Barbara, CA 93106, USA}
\affiliation{$^3$Department of Mathematics, Physics, and Statistics, University of British Columbia, \\ Okanagan Campus, 1177 Research Road, Kelowna, BC V1V 1V7, Canada}

\begin{abstract}
We study how bounds on new $U(1)$ long-range forces, resulting from gauging the Standard Model symmetries $\BL$ and $L_a-L_b$, $a,b=e,\mu,\tau$, are modified if the mediator mass is generated by the Higgs mechanism.
Large baryonic and leptonic charges of astronomical bodies displace the Higgs field from its vacuum value, resulting in charge screening.
Depending on the $U(1)$ Higgs mass,
the screening relaxes the bounds on the new force strength and mediator mass from equivalence principle violation and neutrino oscillations experiments, and eliminates the bounds on the muonic $\Lmt$ force from neutron star binaries.
\end{abstract}

\maketitle


\emph{\textbf{Introduction}} ---
New light $U(1)$ gauge bosons resulting from gauging the global symmetries of the Standard Model (SM) give rise to a variety of observational signatures~\cite{ParticleDataGroup:2022pth}.
They mediate long-range forces that are constrained by 
experiments looking for a fifth force or equivalence principle violation (EPV)~\cite{Antypas:2022asj}.
Within the SM, 
it is natural to consider gauging the following four combinations of baryon and lepton charges: $\BL$, $\Lem$, $\Let$ and $\Lmt$.

Large baryon and electron charges of astronomical bodies, such as the Earth or the Sun, lead to large $\BL$ and $\LLo$ field strengths sourced by the charge which, unless screened, are severely constrained by fifth-force and EPV experiments~\cite{Adelberger:2003zx,Adelberger:2006dh,Schlamminger:2007ht,Wagner:2012ui,MICROSCOPE:2022doy}.
The constraints of the $\Lmt$ force are much weaker and stem from observations of neutron star (NS) binary dynamics and cooling~\cite{KumarPoddar:2019ceq,Dror:2019uea,Fiorillo:2026wso,Liu:2026dbd}.
These constraints are given in the Proca limit, i.e. assuming that the new gauge boson is the only field determining the field strength sourced by a charged body. However, the ultralight vector boson mass $m_V$ is expected to appear dynamically, for example, through the Higgs mechanism~\cite{Reece:2018zvv}. This introduces additional degrees of freedom that can affect the field of the source.

The Abelian Higgs Model (AHM) is a candidate UV completion of the Proca theory, in which $m_V$ is generated by the Higgs vacuum expectation value (vev). In this minimal setting,
endowing the massive gauge boson with the Higgs mechanism introduces one new parameter: the Higgs field self-interaction $\l$ or $\tg\equiv g/\sqrt{\l}$, where $g$ is the gauge coupling.
Crucially, the Higgs field is charged under the new $U(1)$.
It has been known that in abelian gauge theories light charged scalars can screen large charged bodies~\cite{Mandula:1977sq,Mandula:1977gy,Hey:1978bg, Bawin:1983pok,Bawin:1988na, Lee:1955vk,Okun:1969ey, Blinnikov:1995kp,Martemyanov:1997kd,Martemyanov:1998dj}.
This is a vector analog of the well-studied screening of scalar-mediated forces ~\cite{Khoury:2003aq,Hinterbichler:2010es,Khoury:2010xi,Burrage:2016xzz,Olive:2007aj,Gubser:2004uf,Blinov:2018vgc,Brax:2021rwk,Brax:2023qyp,Chauhan:2024qew, Brax:2021wcv}.
Implications of screening for the searches of new vector forces have been first discussed in~\cite{Nelson:2008tn} where the AHM for the $B-L$ force was considered; see also~\cite{Ishihara:2018eah, Forgacs:2020sms}.
In this \emph{Letter} we comprehensively analyze the screening effect in various regimes, depending on the value of $\tg$, and explore its phenomenological consequences for the four different $U(1)$ forces.


\emph{\textbf{Abelian Higgs Model}} ---
The AHM Lagrangian takes the form
\begin{equation}\label{Lagr}
    \mathcal{L}=-\frac{1}{4}F_{\mu\nu}F^{\mu\nu} + |D_\mu\phi|^2 - \frac{\l}{4} (|\phi|^2-v^2)^2 - gA_\mu J^\mu \;.
\end{equation}
Here $A_\mu$ and $\phi$ are the gauge and Higgs fields, respectively, $F_{\mu\nu}=\d_{\mu} A_{\nu} - \d_{\nu} A_{\mu}$, $D_{\mu}=\d_{\mu}-igA_{\mu}$, $J^{\mu}$ is the external current associated with one of the four combinations of baryon and lepton numbers, and we assume that the Higgs field has a unit charge. Let $\phi=v$ be the vacuum state of the Higgs field.
The vacuum masses of the gauge boson and Higgs are $m_V=\sqrt{2}gv$ and $m_\phi=\sqrt{\l}v$. In the limit $m_\phi\to\infty$, $m_V=\rm{const}$ we recover the Proca theory for the gauge boson. We will study deviations from the Proca theory at a finite $m_\phi$ and in the presence of a background charge distribution.
Experimental constraints on $g$ and $m_V$ will be conditional to the value of $\tg=m_V/(\sqrt{2}m_\phi)$.

Consider a static, uniform and spherically-symmetric charge distribution of size $R$, charge density $n$ and total charge $N$. Then $J^\mu=(\rho,0)$ with $\rho=n$ at $r<R$ and $\rho=0$ otherwise.
The fields generated by this distribution depend only on the radial coordinate $r$.
Fixing the gauge as $\vec{A}=0$ leads to the equations of motion for $f\equiv|\phi|$ and $V\equiv A^0$,
\begin{subequations}\label{EOMs}
\begin{align}
    & \Delta f + g^2 V^2 f-\frac{\l}{2}f(f^2-v^2)=0 \;, \label{eqRho}\\
    & (\Delta - 2g^2f^2)V=-g\rho \label{eqA} \;,
\end{align}
\end{subequations}
paired with the vacuum boundary condition at infinity, $f(\infty)=v$, $V(\infty)=0$, and the regularity condition at the origin, $f'(0)=V'(0)=0$, where prime means derivative with respect to $r$.
In the limit $f=v$, \cref{eqA} reduces to the Proca equation for the massive gauge boson. However, $f=v$ is not the solution in the presence of external charges: the Higgs field is displaced from its vev. When the displacement is large enough, it back-reacts on the gauge potential resulting in charge screening. 
The physical reason is that in the presence of large external charge it is energetically favourable to excite a cloud of oppositely-charged particles that screen the Coulomb field around the source. The energy gain from the screening must be larger than the energy cost of having the cloud. The latter can be due to the gradient term~\cite{Mandula:1977sq} or self-interaction potential~\cite{Nelson:2008tn} of the charged scalar.

In general, eqs.~(\ref{EOMs}) must be solved numerically; however, it is useful to study first various asymptotics of the strong-screening solution analytically. 

By \cref{eqRho}, the gauge potential shifts the minimum of the Higgs potential to larger $f$.
The strong screening can be achieved if $f\gg v$.
The ``effective'' masses of the gauge and Higgs fields in this regime are
\begin{equation}\label{EffMasses}
    \mV = \sqrt{2}gf \;, ~~~ \mF= \sqrt{\l}f \;,
\end{equation}
and $\mV/m_V\gg 1$ at strong screening. We will further require $\mV R \gg 1$ which correspond to screening of the source. 
On the other hand, $\mF$ can be both large and small compared to $R^{-1}$. This leads to two screening regimes.

{\bf\emph{Thin-shell screening}} ---
Assume that $\mF R\gg 1$ inside the source. Since also $\mV R\gg 1$, the magnitudes of the both fields stay constant at $r<R$. Solving \cref{eqRho,eqA} for constant functions, we obtain
\begin{equation}\label{StrongScreening}
\bar{V}=\frac{(3N)^{1/3}}{(16\pi\tg^2 g)^{1/3}R}\;, ~~~ \bar{f}=\sqrt{2}\tg \bar{V} \;.
\end{equation}
For this solution the charge density induced by the Higgs field, $q_{\rm ind.}=-2g^2\bar{f}^2\bar{V}$, is exactly opposite to the charge density of the source, hence the charge is totally screened in the bulk of the body. However, the screening is not perfect at the surface of the body. 
The thickness of the surface layer is $\propto \mV^{-1}\ll R$. This ``thin-shell'' screening regime has been discussed before both for scalar- and vector-mediated forces; see, e.g., \cite{Martemyanov:1997kd,Nelson:2008tn}.
Using \cref{StrongScreening}, the condition $\mF R\gg 1$ of validity of the thin-shell regime becomes $N g^2/\tg^2\gg 1$, which holds automatically at strong screening for $\tg < 1$. 

Next, let us evaluate the gauge potential outside the source, $r\geqslant R$. We present simple analytic approximations in the limit $m_VR\ll 1$, postponing the full numerical solution for later. Two cases can be distinguished depending on whether $\mF\gg\mV$ ($\tg\ll 1$) or $\mF\ll \mV$ ($\tg\gg 1$).

In the first case, the Higgs field follows the gauge field profile, $f^2\approx2\tg^2V^2$ 
as long as $f\gg v$. Substituting this to \cref{eqA}, we find
\begin{align}\label{PotSelfFar}
    & V(r)\approx\frac{\bar{V}}{1+\mVbar(r-R)/\sqrt{2}} \;,
\end{align}
where $\mVbar=2^\frac{1}{3}g^\frac{2}{3}\tg^{\frac{1}{3}}n^\frac{1}{3}$.
We see that the surface layer extends to the distance $\mVbar^{-1}$ from the surface of the body.
The solution (\ref{PotSelfFar}) is valid in the planar approximation of the source, $r-R\ll R$, and until the exponential suppression due to the vacuum gauge boson mass kicks in at $r-R\sim m_V^{-1}$.
If $m_V^{-1}\gg R$, then at $r-R\gtrsim R$ the potential
is approximately Coulomb, with the effective charge number
and a logarithmic correction:
\begin{equation}\label{Neff2}
  \hat{N} = \frac{\sqrt{2}\pi}{g^2 \tg} \;, ~~ V(r)\approx \frac{g \hat{N}}{4\pi r\sqrt{\log (r/R)}} \;.
\end{equation}
We will neglect this correction in the analytic estimates below, except when discussing the $\Lmt$ force.

In the light Higgs case, the Higgs field outside the source varies slower than the gauge field, hence 
$\mV\approx\mVbar$
and the solution reads
\begin{equation}\label{ForceSelf1}
\begin{split}
    V(r)\approx gnR\frac{\e^{-\mVbar (r-R)}}{\mVbar^2 r} \,.
\end{split}
\end{equation}
It matches \cref{StrongScreening} at $r=R$ and is valid
until $r-R\sim \mF^{-1}$ where it is exponentially suppressed. 
Crucially, the force range is determined by $\mV$ which depends on $g$. Increasing the gauge coupling shortens the range, making EPV experiments insensitive to the new force at sufficiently large coupling.

\textit{\textbf{Gradient screening}} --- 
Assume that $\mF R\ll 1$ inside the source. Then the interaction term between the Higgs and gauge bosons in \cref{eqRho} is balanced not by the Higgs self-interaction but by the gradient term, 
analogously to the massless scalar screening the Coulomb charge~\cite{Mandula:1977sq}.
There is no exact charge density cancellation inside the body, but screening can still be efficient, as we will see.
Since $\mV\gg \mF$ at $r<R$, the gauge potential is $V(r)\approx gn/\mV^2$.
Substituting this to \cref{eqRho} and assuming parametrically $\Delta f\sim \bar{f}/R^2$, we obtain
\begin{equation}\label{GradScreening}
    \bar{f}\sim \frac{N^{1/2}}{R} \;, ~~ \bar{V}\sim \frac{1}{g R} \;.
\end{equation}
From here and \cref{EffMasses} we obtain that inside the source $\mF R\sim g N^{1/2}\tg^{-1}$. Since $\mV R\sim gN^{1/2}$, the gradient regime requires $\tg\gg 1$. Furthermore, requiring the gradient term in \cref{eqRho} dominate over the self-interaction term gives $N g^2/\tg^2\ll 1$, which is opposite to the validity condition of the thin-shell regime.
Crucially, the charge density induced by the Higgs field is parametrically the same as the charge density of the source,
$q_{\rm{ind.}}\sim g^2 \bar{f}^2\bar{V} \sim -gn$, 
showing that a strong screening can be achieved in the gradient regime. 

Outside the source, $V$ drops faster than $f$ given that $\mV\gg \mF$. 
Hence, at $r-R\ll \mF^{-1}$ the Higgs field does not change much from its surface value $\bar{f}$ and $V$ is given by \cref{ForceSelf1}.

\emph{\textbf{\boldmath{$B-L$}}} --- 
The strongest bounds on the long-range $B-L$ force come from MICROSCOPE~\cite{MICROSCOPE:2022doy,Fayet:2025brx} and E\"{o}t-Wash~\cite{Schlamminger:2007ht,Wagner:2012ui}.
In MICROSCOPE the source of the force is the Earth itself.
It is convenient to define the parameter $\geff$ equal to the experimental upper bound on $g$ in the Proca theory. To translate this bound to the bound on $g$ in the AHM, one should solve the equation
\begin{equation}\label{EqForGEff}
   F(r)\equiv -g V^\prime(r)\le F_\mathrm{lim} \approx \frac{\geff^2N}{4\pi R^2} \;,
\end{equation}
where the force $F(r)$ is calculated at the distance from the center of the body at which an experiment is performed, and we assumed $r\approx R$.
We approximate the Earth by a uniformly charged spherically-symmetric body with $R= 6300$~km and the neutron number $N=2\cdot 10^{51}$. MICROSCOPE's orbit is $L \approx 700$ km from the surface. 
To showcase how the Proca bound changes, we assume $m_VR\ll 1$ or $m_V\lesssim 10^{-13}$~eV in the parametric estimates below.
For this mass, the MICROSCOPE bound is $\geff\simeq 10^{-25}$~\cite{MICROSCOPE:2022doy,Fayet:2025brx}.

The thin-shell screening is achieved at $\mV R\gtrsim 1$ or $\tg g^2N\gtrsim 1$. Since $g\gtrsim \geff$, this implies $\tg \gtrsim 10^{-1}$. 
Assume that $\mV L \ll 1$. 
Then from \cref{PotSelfFar,EqForGEff} we obtain $g \sim N^{1/4} g_\mathrm{eff}^{3/2} \tg^{1/4}$.
When $\tg$ increases above $\tg_* \equiv g  N^{1/2}\sim 10$, the thin-shell regime changes to the gradient one. 
Crucially, the force is $\tg$-independent in the gradient regime and the screening effect asymptotes.
Thus, the upper bound on $g$ is relaxed at most by $g/g_\mathrm{eff} \sim N^{1/4} g_\mathrm{eff}^{1/2} \tg_*^{1/4} = \mathcal{O}(1)$.
We conclude that the Higgs screening does not significantly modify the MICROSCOPE's upper bound on the gauge coupling at all values of $\tg$.

At the same time, for $\tg \gtrsim 1$ increasing $g$ decreases the range of the force, and at $\mV L>1$ the constraint disappears due to the exponential screening of the force at the satellite location.
This imposes the \textit{lower} bound on $g$ and opens the range $g\gtrsim 10^{-24}$ in the gradient regime $\tg\gtrsim 10$.

The region of $g,~m_V$ unconstrained by MICROSCOPE at strong screening is partially covered by the E\"{o}t-Wash experiment, which is sensitive to the horizontal component of the force from different sources, ranging from local environment to the whole Earth depending on the value of $m_V$.
For $m_V\lesssim 10^{-13}$~eV, the E\"{o}t-Wash experiment feels the entire Earth's force and constrains $\geff\simeq 10^{-24}$~\cite{Wagner:2012ui}.
Repeating the above argument, the screening affects the bound at $\tg \gtrsim 10^{-3}$. At $\tg_*\sim 100$ the screening enters the gradient regime, and the bound saturates at $g/\geff=\mathcal{O}(10)$. 
Thus, the range $g\gtrsim 10^{-23}$ is excluded by E\"{o}t-Wash regardless of the value of $\tg$.

\begin{figure}
    \centering
    \includegraphics[width=0.99\linewidth]{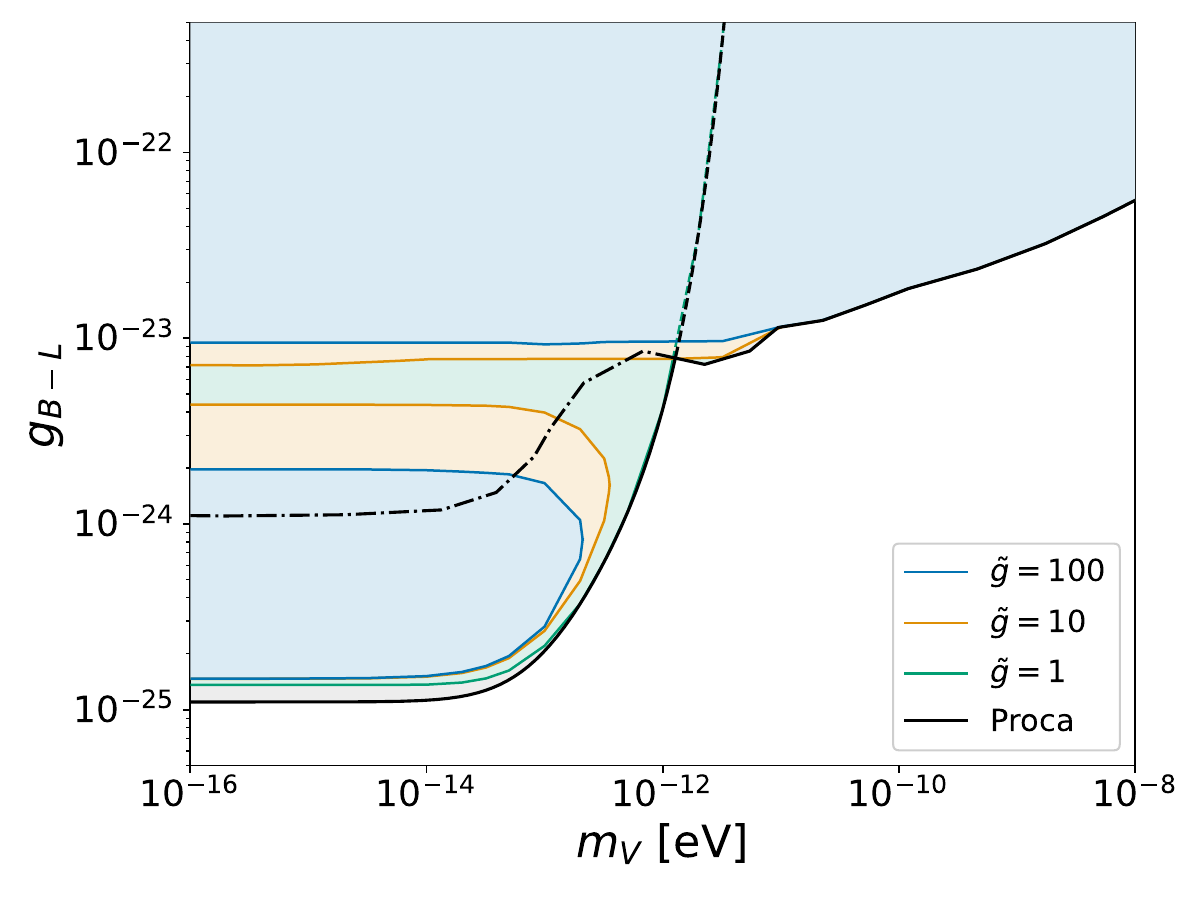}
    \caption{Exclusion regions on the $\BL$ force in the AHM at different values of $\tg$. The MICROSCOPE and E\"{o}t-Wash bounds are combined. The dashed (dash-dotted) lines show sub-leading MICROSCOPE (E\"{o}t-Wash) bounds.}
    \label{fig:micro_ew}
\end{figure}

This analysis is confirmed by computing the horizontal force at the experiment location from the exact gauge potential which we find numerically. The combined bounds on the $\BL$ coupling at several values of $\tg$ are shown in Fig.~\ref{fig:micro_ew}. We see the window $10^{-24}\lesssim g\lesssim 10^{-23}$ that is unconstrained by either experiment in the strong-screening regime and $\tg\gg 1$.

At $m_V >10^{-13}$~eV E\"{o}t-Wash is sensitive to the local field sources. Their effect can be evaluated by solving for the field perturbations in the background Earth fields.
In the gradient regime $\tg \gg 1$ and at values of $g$ comparable to $\geff$, the effective mass $\mV$ is approximately constant on distances between the experiment and the local sources.
Hence the bound on $g$ at a given $m_V >10^{-13}$~eV equals the Proca theory bound $\geff$ at the corresponding value of $\mV$.
On the other hand, in the thin-shell regime the scale of variation of the background potential and the effective gauge boson mass are both set by $\mVbar$. When $\mVbar^{-1}\ll R$ (corresponding to $0.1\lesssim \tg\lesssim 10$ at $g\sim\geff$),  
a dedicated analysis of local source screening is needed to find how the bound on $g$ changes, hence we do not show these bounds on Fig.~\ref{fig:micro_ew}. Furthermore, similarly to MICROSCOPE, in the gradient regime the E\"{o}t-Wash bounds are two-sided. In particular, the R\"{o}t-Wash~\cite{Smith:1999cr} bound disappears at $g\gtrsim 10^{-17}$ when the force range becomes smaller than the distance to the attractor $\approx 7$~cm.
To properly find the upper boundary of the exclusion region requires the dedicated analysis of local source screening, which is beyond the scope of this work.

\emph{\textbf{\boldmath{$\LLo$}}} --- 
The EPV constraints discussed above apply straightforwardly to the flavourful $\LLo$ forces~\cite{Wagner:2012ui,MICROSCOPE:2022doy,Fayet:2025brx}.
Another type of constraints on these forces comes from neutrino oscillations. We assume that active neutrino species are coupled to the new gauge boson via the vector current as in \cref{Lagr}. Neutrino traversing the Earth or the Sun oscillate in the flavour-electric potential generated coherently by electrons, which contributes to the Hamiltonian of neutrino oscillations similarly to the local matter Hamiltonian~\cite{ParticleDataGroup:2024cfk}. At low gauge boson masses, $m_V\lesssim 10^{-13}$~eV, the matter effect induced by the new potential can be a more sensitive probe of the $\LLo$ forces than the fifth-force tests~\cite{Grifols:2003gy,Joshipura:2003jh}. 

\begin{figure}
    \centering
    \includegraphics[width=0.99\linewidth]{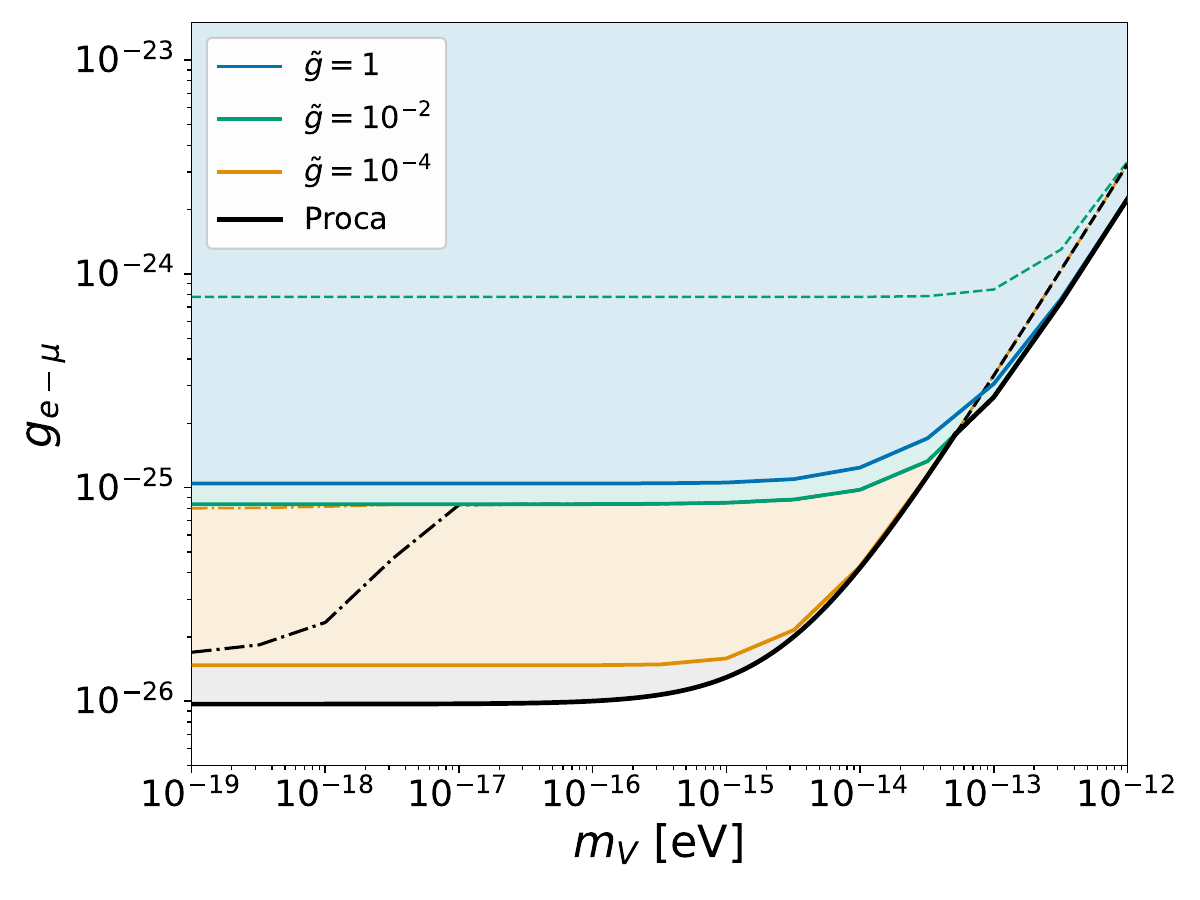}
    \caption{Constraints on the $\Lem$ force in the AHM at several values of $\tg$ from solar and terrestrial neutrino data. The dashed (dash-dotted) line shows the subleading solar (terrestrial) neutrino bounds.}
    \label{fig:em}
\end{figure}

We need to distinguish two types of neutrino constraints, the one using solar neutrino data~\cite{Minakata:2010be,Coloma:2020gfv} and another using terrestrial neutrino sources~\cite{Wise:2018rnb,Coloma:2020gfv}. The reason is that the Sun and the Earth enter the strong-screening regime at different values of $\tg$. Furthermore, the solar flavour-electric potential can dominate over the terrestrial potential at the Earth's orbit if $m_V<1/$A.U.~$\sim 10^{-18}$~eV.

For solar neutrinos, to a good approximation we can assume the potential $V_\odot$ be given by its value at the origin $V_\odot(0)$, since the neutrino production takes place close to the center of the Sun~\cite{Coloma:2020gfv}. Similarly, we evaluate the terrestrial potential $V_\oplus$ near the surface~\cite{Wise:2018rnb}.
The experimental bounds on the extra solar~\cite{Minakata:2010be,Coloma:2020gfv} and terrestrial~\cite{Super-Kamiokande:2011dam, Wise:2018rnb} matter potential energies are:
\begin{equation}\label{VMax}
    U_{\odot,\rm{max}} \approx 1\cdot10^{-11} \,\text{eV} \;, ~~~ U_{\oplus,\rm{max}}\approx 3\cdot 10^{-14} \,\text{eV} \;.
\end{equation}
In the Proca theory, they result in the bounds $g\lesssim 10^{-25}$ for terrestrial neutrinos in the Earth's potential ($m_V\lesssim 10^{-13}$~eV); $g\lesssim 10^{-26}$ for terrestrial neutrinos in the solar potential (as long as the potential reaches the Earth's orbit, $m_V\lesssim 10^{-18}$~eV); and $g\lesssim 10^{-26}$ for solar neutrinos (at $m_V\lesssim 10^{-15}$~eV). To translate these bounds to the bounds on $g$ in the AHM, we need to solve $gV= U_{\rm max}$
where, depending on the type of the bound, $V$ is identified with $V_\oplus$, $V_\odot(0)$ or $V_\odot(L_\oplus)$, and $L_\oplus=1.5\cdot 10^8$~km is the radius of the Earth's orbit. 

First, consider the solar neutrino bound. In the thin-shell regime 
\cref{StrongScreening} gives $gV_\odot(0)\sim n(0)^{\frac{1}{3}}g^{\frac{2}{3}}\tg^{-\frac{2}{3}}$, where 
$n(0)\approx 1.5\cdot 10^{26}$~cm$^{-3}$
is the electron number density in the center of the Sun. Equating this to $E_{\odot,\rm{max}}$ we find that the solar-neutrino bound on $g$ is modified at $\tg\gtrsim 10^{-4}$. 
Furthermore, since $g\propto\tg$, at $\tg\gtrsim 10^{-3}$ the solar-neutrino bound is weaker than the terrestrial-neutrino bound.
Next, consider the fate of the solar potential at the Earth's orbit. Using the Coulomb solution (\ref{PotSelfFar}) we find that $gV_\odot(L_\oplus)\sim 1/(\tg L_\oplus)$ and is independent of $g$. Equating this to $U_{\oplus,\rm{max}}$ gives $\tg\sim 10^{-4}$. We conclude that at $\tg\gtrsim 10^{-4}$ the terrestrial-neutrino bound using the solar potential disappears.
Finally, consider the terrestrial flavour-electric potential. Proceeding as above, we find that the bound on $g$ is affected by screening at $\tg\gtrsim 10^{-1}$.
When the solution enters the gradient regime at $\tg\gtrsim 1$, the bound asymptotes to a constant value where it is comparable to the EPV bounds.

We confirm this analysis by using the full numerical solution of the screened gauge potential. 
We use the exponential electron density profile in the Sun to compute the solar potential, $n(r)=n(0)\exp(-r/r_*)$ with $r_*=6.6\cdot 10^4$~km~\cite{Bahcall:1987jc}.
The resulting constraints for $\Lem$
are presented in Fig.~\ref{fig:em};
similar bounds hold for $\Let$~\cite{Coloma:2020gfv}.

\emph{\textbf{\boldmath{$\Lmt$}}} --- 
The strongest bounds on $\Lmt$ come from NS binary dynamics. This is because NSs contain a significant ($\sim0.1-1\%$) mass fraction of
muons~\cite{haensel2007neutron}.
Binary systems with a NS form dipoles and experience 
an extra energy loss through $\Lmt$ dipole boson radiation~\cite{Dror:2019uea}. This effect sets leading constraints on the gauge coupling for $m_V \lesssim \Omega$ where $\Omega$ is the angular frequency of the binary which for simplicity we assume to be circular; see~\cite{Krause:1994ar} for the general treatment.
By \cref{StrongScreening}, the NS is screened if $\tg \gg  1/(g^2 N)$, and its effective charge is given in \cref{Neff2}. The unscreened charge is distributed in a shell of the size comparable to the radius of the NS, which is much smaller than the distance between the bodies. Using \cref{Neff2}, the dipole moment of the system at strong screening is estimated as
\begin{equation}\label{eq:higgs_dipole}
    d \simeq\frac{\pi a}{\tg g {\sqrt{\log 1/(\Omega R)}}}\frac{M_1-M_2}{M_1+M_2} \;,
\end{equation}
where $a$ is the orbit semimajor axis and $M_{1,2}$ are masses of the two bodies.
Thus, unlike the Proca limit, at strong screening the dipole radiation power decreases as $g$ increases.
As a result, there exists a binary-dependent critical value $\tg_\mathrm{crit}$ such that at $\tg > \tg_\mathrm{crit}$ the dipole radiation bound on $g$ for a given system disappear. Comparing with the expression for $d$ in absence of screening, we obtain $\tg_\mathrm{crit} \sim (M_2-M_1)/(\geff^2(N_1M_2-N_2M_1))$, where $N_{1,2}$ are the total muonic charges and $\geff$ is the experimental bound on $g$ in the Proca theory.

\begin{figure}
    \centering
    \includegraphics[width=0.99\linewidth]{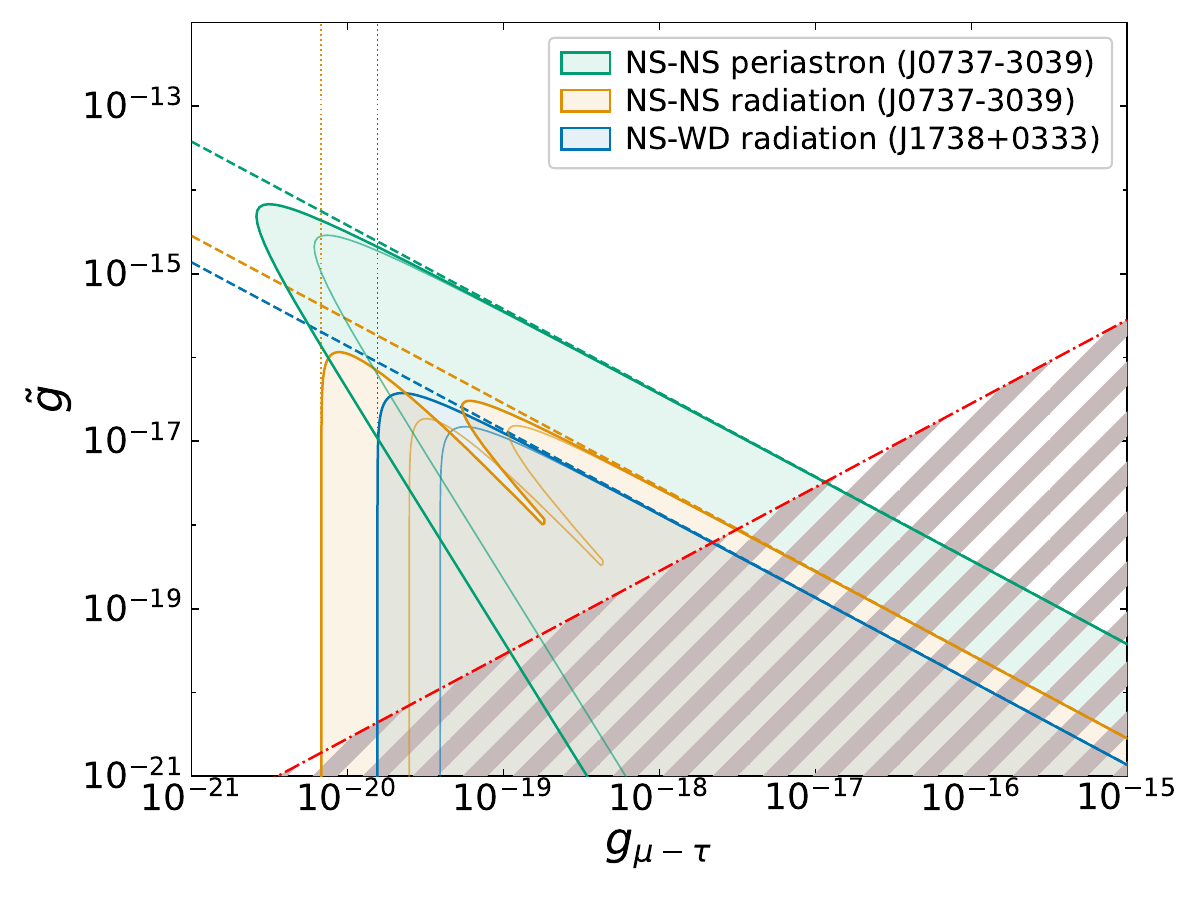}
    \caption{Bounds on the $\Lmt$ force in the AHM from NS-NS binary J0737-3039 and NS-WD binary J1738+0333 for $m_V\ll 10^{-19}$~eV and two choices of the NS EoS~\cite{Pearson:2018tkr,Dror:2019uea}. The dashed lines show the slopes $\tg\propto g^{-1}$ predicted by eqs.~(\ref{eq:higgs_dipole}), (\ref{Periastr2}). The dotted lines show the bounds on $g$ in the Proca limit~\cite{Liu:2025zuz}. 
    The hatched region is excluded by unitarity.
    }  
    \label{fig:Lmt_dipole_bound}
\end{figure}

The $(g,\tg)$ exclusion region is shown in Fig.~\ref{fig:Lmt_dipole_bound} using the double pulsar J0737-3039~\cite{PhysRevX.11.041050} which gives the strongest constraint and assuming $m_V\ll \Omega\sim 10^{-19}$~eV.
We solve \cref{EOMs} numerically to find the effective charges of the NSs; the result agrees well with the approximation in \cref{eq:higgs_dipole}. 
We assumed a uniform muon distribution in the NSs with the optimistic (pessimistic) muon mass fraction calculated using BSk22 (BSk26) NS EoS~\cite{Pearson:2018tkr,Dror:2019uea}.
The Higgs cloud enveloping the binary was approximated by a superposition of the individual clouds; we checked that nonlinear corrections do not change the result significantly.
Note that typically the two NSs enter the strong screening at different $\tg$. Let star 1 enter it first, then at $\tg\sim M_2/(\geff^2N_2M_1)$ the dipole moment of the system vanishes and the constraint disappears.
It reappears at larger $\tg$ as $d$ keeps decreasing, producing a ``notch'' in the $(g,\tg)$ exclusion region shown 
in Fig.~\ref{fig:Lmt_dipole_bound}.
For completeness, we also show the exclusion region from a NS-WD binary J1738+0333~\cite{Antoniadis, Freire:2012mg} which has a convex shape.

The second type of bounds on the $\Lmt$ force is 
associated with the periastron advance caused by deviations from the inverse square law force~\cite{Liu:2025zuz}.
Assuming an extra interaction energy 
of the form $U_{12} = \alpha g_{12} u(r)/r$, $g_{12}\equiv G_NM_1M_2$, the periastron advance in the orbit of major semiaxis $a$ at small eccentricity is~\cite{Adkins:2007et}
\begin{equation}\label{eq:periastron_advance}
    \Delta \varpi = - \alpha a^2 \pi u^{\prime\prime}(a) \;.
\end{equation}
In the Proca theory, the deviation from the inverse square law is due to the Yukawa suppression of the force and is significant at $m_Va\gtrsim 1$.
The presence of the Higgs modifies the interaction potential even at $m_Va\ll 1$. Furthermore, the deviation becomes significant before the strong screening is reached. To see this, we note that in the limit $\tg\ll 1$ the Higgs field outside the NS behaves as $f^2\approx v^2+2\tg^2V_0^2$ where $V_0$ is the NS Coulomb potential. Substituting this to \cref{eqA} and expanding $V$ in powers of $\tg^2$, we find the correction to the Coulomb potential $\delta V\simeq -\tg^2g^5N^3\log(m_Vr)/(16\pi^3r)$ valid at small $m_Vr$. From here to the leading order in $\tg^2$ we obtain
\begin{equation}\label{Periastr1}
    u(r)\simeq-\gamma_{12} \log(m_Vr) ~\Rightarrow ~ \Delta\varpi \simeq - \frac{g^2N_1N_2\gamma_{12}}{64\pi^2g_{12}} \;,
\end{equation}
where $\gamma_{12}\equiv4\tg^2g^4(N_1^2+N_2^2)$. 
On the other hand, in the strong-screening regime the NS potential is given in \cref{Neff2}.
Assuming for simplicity that both stars have the same size $R$, in the limit $\log (r/R) \gg 1$ we obtain
\begin{equation}\label{Periastr2}
   u(r)\simeq \frac{\pi}{g^2\tg^2\log (\frac{r}{R})} ~\Rightarrow ~\Delta\varpi \simeq -\frac{\pi^{2}}{g_{12}g^2 \tg^2 \log^{2} (\frac{a}{R})} \;.
\end{equation}
From \cref{Periastr1,Periastr2} we see that the $(g,\tg)$ exclusion region is a wedge formed by the lines $\tg\propto g^{-3}$ and $\tg\propto g^{-1}$. This is confirmed by using the exact numerical solution for the NS potential energy. The periastron bounds from the double pulsar J0737-3039 are shown in Fig.~\ref{fig:Lmt_dipole_bound}.
We see that there is no bound on $g$ at $\tg\gtrsim 10^{-14}$.
Similarly, for $m_Va\gtrsim 1$ the bound disappears at $\tg\gtrsim 10^{-13}$.
We conclude that Higgsing the long-range muonic force removes the bound on $g$ at almost all values of $\tg$ allowed by unitarity, $\tg>g/\sqrt{4\pi}$.

In this analysis we have neglected back-reaction of the gauge and Higgs fields on the NS EoS. The latter is important when the $\Lmt$ potential energy in the NS is comparable to the muon mass, 
$U\sim 100$ MeV. In the Proca limit, this happens when $g$ reaches $10^{-18}$~\cite{Dror:2019uea,Liu:2025zuz,Liu:2026dbd}.
Assuming no direct coupling of the Higgs to the NS matter, the screening delays the onset of back-reaction by suppressing the gauge potential. Solving $g\bar{V}\lesssim U$, with $\bar{V}$ given in \cref{StrongScreening}, gives $g\lesssim\max(10^{-18},10\,\tg)$,
which is always satisfied in the perturbative AHM.

\emph{\textbf{Discussion}} --- 
In this \emph{Letter} we discussed how the Higgs mechanism for abelian gauge bosons impacts experimental constraints on new long-range vector forces.
First, we revisited bounds on the $\BL$ force from MICROSCOPE and E\"{o}t-Wash experiments~\cite{MICROSCOPE:2022doy,Schlamminger:2007ht,Wagner:2012ui}. We found that the upper bound $g_{B-L}\lesssim 10^{-25}$ is surprisingly robust 
and does not depend on $\tg=m_V/(\sqrt{2}m_\phi)$. At the same time, the excited Higgs field reduces the force range and makes MICROSCOPE insensitive to couplings $g\gtrsim 10^{-24}$ at $\tg\gg 1$, similarly to the chameleon mechanism for scalar forces~\cite{Khoury:2003aq}. The E\"{o}t-Wash covers partially the reopened region; nevertheless, a part of the $(m_V,g)$ parameter space remains open at $\tg\gg 1$.

We then turned to the flavourful $\LLo$ forces which can be probed with neutrino oscillations.
We found that the strongest upper bound $g_{e-\mu/\tau}\lesssim 10^{-26}$ coming from solar neutrino data at $m_V\lesssim 10^{-15}$~eV~\cite{Minakata:2010be,Coloma:2020gfv} is alleviated due to the Higgs screening the charge of the Sun as soon as $\tg\gtrsim 10^{-4}$. The leading constraint $g_{e-\mu/\tau}\lesssim 10^{-25}$ is then provided by terrestrial neutrino experiments~\cite{Wise:2018rnb}. At $\tg\gtrsim 10^{-1}$ the Higgs screening the Earth's charge begins affecting this constraint, and at $\tg\sim 1$ the latter becomes comparable to the MICROSCOPE bound.

Finally, we studied the screening of the $\Lmt$ force in NS binaries. We found that the dipole radiation~\cite{Dror:2019uea} (periastron advance~\cite{Liu:2025zuz}) constraints on $g_{\mu-\tau}$ disappear at $\tg\gtrsim 10^{-16}$ ($\tg\gtrsim 10^{-13}$). Furthermore, the screening causes deviations from the inverse square law of the $\Lmt$ force contributing to the periastron advance of binary's orbit~\cite{Liu:2025zuz}. This \textit{strengthens} the constraint on $g_{\mu-\tau}$ at $m_V\lesssim 10^{-19}$~eV and certain values of $\tg$. 

Our findings can be developed in several directions. 
First, for torsion balance experiments, screening the local environment could affect the bounds already at $\tg<1$. Since in this regime the effective gauge boson mass changes on a scale comparable to distances between local masses and experiment, one cannot approximate the local potential by Yukawa potentials, and a full numerical solution is needed.
Next, we neglected possible interactions between neutrino and the new $U(1)$ Higgs field. The Higgs can play a role in the neutrino mass generation, and this can impact the bounds~\cite{Dror:2020fbh}, albeit in a model-dependent way. 
Third, it would be interesting to study effects due to large $U(1)$ magnetic fields which in the AHM can lead to formation of vortices.
Magnetic effects in the AHM framework have been explored for the SM electromagnetic photon~\cite{Adelberger:2003qx,Dvali:2025mqr,Gruzinov:2026thy} and dark photon~\cite{East:2022rsi}. 
Finally, the influence of the Higgs field on the dynamics of NS inspirals and ensuing gravitational-wave signal would be interesting to investigate, since this gives the strongest projected bounds on the muonic force~\cite{Dror:2019uea}.

\section*{Acknowledgments}

We thank Asimina Arvanitaki, Anatoli Fedynitch, Junwu Huang, Maxim Pospelov, Mikhail Shaposhnikov, Sergey Sibiryakov for useful discussions. Throughout the work Opus 5.5, Fable 5, 5.2 were used via Claude Code, along with GPT 6.2, for writing numerical codes, literature review, checking formulas and proofreading. 
Research at Perimeter Institute is supported in part by the Government of Canada through the Department of Innovation, Science and Economic Development Canada and by the Province of Ontario through the Ministry of Colleges and Universities. 
This research was supported in part by grant NSF PHY-2309135 to the Kavli Institute for Theoretical Physics (KITP), and has benefited from the Mitchel Postdoctoral Scholar Career Development Fund.

\bibliography{Refs}

\end{document}